\documentclass[twocolumn,prl]{revtex4}
\newcommand{\figurewidth}{84mm}

\usepackage{graphicx}
\usepackage{dcolumn}
\usepackage{amsmath}

\begin{document}

\newcommand{\leqs}{\stackrel{\scriptstyle<}{\scriptscriptstyle\sim}\:}
\newcommand{\geqs}{\stackrel{\scriptstyle>}{\scriptscriptstyle\sim}\:}
\newcommand{\PP}{{\mathcal{P}}}
\newcommand{\NN}{{\mathcal{N}}}
\newcommand{\NP}{N_{\cal P}}
\newcommand{\KP}{K_{\cal P}}
\newcommand{\UP}{U_{\cal P}}
\newcommand{\PR}{P}
\newcommand{\RR}{{\bf R}}
\newcommand{\RRp}{{\bf R'}}
\newcommand{\kk}{\mathbf{k}}
\newcommand{\rr}{{\bf r}}
\newcommand{\pp}{\mathbf{p}}
\newcommand{\xx}{\mathbf{x}}
\newcommand{\mm}{{\bf m}}
\newcommand{\dd}{{\bf d}}
\newcommand{\sss}{{\bf s}}
\newcommand{\HH}{{\mathcal{H}}}
\newcommand{\VV}{\mathcal{V}}
\newcommand{\TT}{\mathcal{T}}
\newcommand{\ra}{\rightarrow}
\newcommand{\BE}{\begin{equation}}
\newcommand{\EE}{\end{equation}}
\newcommand{\BEN}{\begin{eqnarray}}
\newcommand{\erf}{\mbox{erf}}
\newcommand{\erfc}{\mbox{erfc}}
\newcommand{\Tr}{\mbox{Tr}}
\newcommand{\Th}{\theta}
\newcommand{\wt}{\tilde{w}}
\newcommand{\kms}{\,\rm km\,s^{-1}}
\newcommand{\gcc}{\,\rm g\,cm^{-3}}
\newcommand{\eV}{\,\rm eV}
\newcommand{\Al}{{\rm Al}}
\newcommand{\p}{{\rm p}}
\newcommand{\s}{{\rm s}}
\newcommand{\K}{{\, \rm K}}
\newcommand{\etal}{{\em et al. }}
\newcommand{\tv}{\tau_{\rm v}} 
\title{
{\sf \vspace*{-12mm}\normalsize Submitted to Physical Review Letters, February 2003.}\\[3mm]
Structure and bonding of dense liquid oxygen
from first principles simulations
}

\author{Burkhard Militzer}
\affiliation{Lawrence Livermore National Laboratory, Livermore, CA 94550}
\author{Fran\c{c}ois Gygi} 
\affiliation{Lawrence Livermore National Laboratory, Livermore, CA 94550}
\author{Giulia Galli}
\affiliation{Lawrence Livermore National Laboratory, Livermore, CA 94550}
\date{\today}
 
\begin{abstract}
Using first principles simulations we have investigated the structural
and bonding properties of dense fluid oxygen up to 180 GPa. We have
found that band gap closure occurs in the molecular liquid, with a
``slow'' transition from a semi-conducting to a poor metallic state
occurring over a wide pressure range. At approximately 80 GPa,
molecular dissociation is observed in the metallic fluid. Spin
fluctuations play a key role in determining the electronic structure
of the low pressure fluid, while they are suppressed at high pressure.
\end{abstract}

\maketitle

The study of pressure-induced transformations in solids and liquids
has become an active field of research in the last
decade~\cite{GeneralReview}, due to key progress both in experimental
methods (diamond anvil cell (DAC) and shock wave experiments) and in
accurate computational tools. In particular, molecular systems such as
hydrogen~\cite{Review_H}, oxygen~\cite{Review_O} and
nitrogen~\cite{Review_N} have been investigated with a variety of
techniques. Nevertheless, fundamental questions on the nature of these
systems in e.g. the dense liquid state are yet unanswered.

In this Letter, we focus on the liquid state of oxygen. In particular,
we examine the structural and bonding changes induced by pressure, as well as the
mechanism leading to the insulator-to-metal transition (ITMT) observed
experimentally in shock wave experiments~\cite{Bastea01}.

Metallization of oxygen was first observed in the solid state by
Desgreniers \etal \cite{Desgreniers} who reported a Drude-type
metallic behavior of its reflectivity for pressures (P) above 96 GPa.
These results were later confirmed by Shimizu and
coworkers~\cite{shimizu95} who observed a change in the sign of the
resistance-temperature curve at 96 GPa, and also discovered
superconductivity in solid, dense oxygen~\cite{shimizu98}. Following measurements by
Akahama~\etal \cite{Akahama}, Weck~\etal \cite{Weck02} showed that in
solid oxygen the ITMT is accompanied by a structural change between
two different crystallographic arrangements of molecules. The
insulating phase below 96 GPa was also studied with
infrared absorption spectroscopy by Gorelli \etal
\cite{Gorelli} who suggested the existence of a lattice of O$_4$
molecules at high pressure. However, this hypothesis is not confirmed
by recent {\em ab initio} calculations~\cite{NA02}, showing that a
geometrical configuration built from extended ``herringbone''-type chains
is energetically favored over a lattice of O$_4$ molecules. The
structure predicted theoretically is consistent with earlier
suggestions based on DAC experiments~\cite{Agnew87}.

In fluid oxygen, metallization was recently reported by Bastea \etal
\cite{Bastea01} who used reverberation shock wave techniques to
generate compression states on a quasi-isentrope. For a given shock speed, 
the thermodynamic states reached in these experiments are at
higher pressure and lower temperatures ($T$) than Hugoniot states
obtained in earlier shock wave measurements~\cite{Ha88}. When
increasing P from $\simeq$ 30 to 120 GPa (and
correspondingly $T$ by several thousands K), Bastea \etal
\cite{Bastea01} observed an increase of the electrical conductivity
over six orders of magnitude. At about 120 GPa and 3.4-fold
compression, the measured conductivity attains values characteristic
of a poor metal. As $P$ is further increased from 120 to 190 GPa, the
conductivity levels off and is nearly constant as a function of
$P$. These experimental results, which indicate that the ITMT occurs
at higher P in the liquid than in the solid, are rather
surprising. In general, one would expect disorder or dissociation
effects to lead to a lower metallization pressure in the fluid than in
the ordered solid.

In order to investigate the mechanism by which the ITMT occurs in the
fluid, we have carried out a series of {\em ab initio} molecular
dynamics (MD) simulations~\cite{CP85} in the liquid state for
different pressures. Our findings shed light on the microscopic
structure and electronic properties of the dense fluid state and on
the key role of spin fluctuations present in fluid before
metallization occurs. Our predicted metallization pressure in the
fluid is lower than that reported experimentally~\cite{Bastea01}, as
well as lower than the metallization pressure measured in the
solid. Our findings further indicate that the system becomes metallic
in a predominantly molecular state, as opposed to other fluids like
hydrogen~\cite{HG02} and nitrogen~\cite{mazevet03}.

We have carried out {\em ab initio} MD simulations~\cite{CP85} using a
gradient corrected energy functional~\cite{PBE} in the density
($\rho$) and temperature range explored in recent shock wave
experiments: $2.8 \leq \rho \leq 4.5 \gcc$ and $1000 \leq T \leq 6400
\K$. In some of our simulations we have treated the spin variable
explicitly within the gradient corrected local spin density
approximation (GCSDA). We performed constant volume microcanonical
simulations~\cite{Jeep} with cells containing 54 and 108 oxygen
atoms. For each $\rho$ and $T$ the total simulation time was
4--5 ps, after equilibration.

\begin{figure}[htb]
\includegraphics[angle=0,width=\figurewidth]{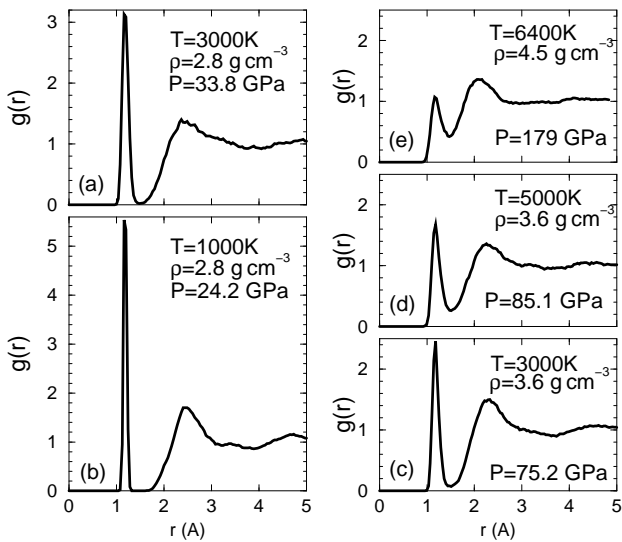}
\caption{
         Pair correlation functions calculated using GCSDA (see text)
         for different densities, temperatures, and pressures as
         specified in the legends.}
\label{gofr}
\end{figure}

Our calculated pair correlations functions ($g(r)$) are shown in
Fig.~\ref{gofr}. Finite size effects were found to be negligible
compared to temperature and pressure effects. At $\rho=2.8\gcc$ and
$T=1000\K$~\cite{weck_melting}, the first sharp peak of $g(r)$ at 1.20$\,$\AA, followed by
a minimum very close to zero, is indicative of a molecular fluid. The
computed pressure is 24 GPa.  The first maximum is at a distance very
close to the bond length of an oxygen molecule in the gas phase in its
triplet ground state (1.207 \AA)~\cite{PBE_molecules}. As the
temperature is increased to 3000$\K$ (and correspondingly the pressure
to 34 GPa) the first molecular peak is broadened while its position
remains unchanged. This trend continues as the fluid is compressed to
$\rho$ = 3.6$\gcc$ at T=3000$\K$ ($P$=84 GPa). Upon further
compression to $\rho$ = 4.5$\gcc$ at T=6400$\K$ (P=179 GPa), the first
molecular peak becomes less intense and the second broad peak moves to
shorter distances. It is interesting to note that the position of the
intramolecular peak remains constant over a pressure range of 150 GPa
and does not shift to larger separations as e.g. in dense liquid
hydrogen~\cite{Ga99}.

The ground state of the O$_2$ molecule is a triplet, and this has been
shown to give rise to interesting structural effects in the low
density fluid~\cite{Pasquarello02}. It is therefore interesting to
explore the effect of explicit spin treatment on the structural
properties of the high pressure fluid as well. We found that spin
effects are negligible on average structural properties both at low
and high pressure, consistent with recent {\em ab initio} calculations
for solid oxygen~\cite{NA02} in the same pressure range. For example
in Fig.~\ref{gofr}, $g(r)$ calculated with and without explicit
treatment of spin are found to be indistinguishable. However, spin
effects are key in determining the optical excitation spectrum of the
low-P fluid, as we will discuss below when we present our results for
the electronic conductivity.

The microscopic structure of the liquid can be further examined by a
cluster analysis. We define a set of atoms as belonging to a given
cluster if these atoms are separated by less than a given cut-off
distance $r_c$. (We chose $r_c=1.50$ \AA, which corresponds to the
first $g(r)$ minimum in the molecular regime.) We then define cluster
lifetimes by simply computing the duration for which a given cluster
remains intact. This analysis yields a very simple picture for our
simulations at low density (2.8$\gcc$) and 1000 and 3000$\K$: we find
that the liquid consists entirely of O$_2$ molecules. A density
increase to 3.6$\gcc$ at 3000$\K$ leads to a small fraction of
dissociated molecules. Although the liquid is still predominantly
molecular (94\% O$_2$), we find 1\% single atoms and 5\% O$_3$-like
clusters. At this density, the average lifetime of O$_2$ molecules is
about 15 vibrational periods of the isolated O$_2$ molecule
(21.1 fs), while O$_1$ and O$_3$ exist only for about one
molecular vibration.
When the temperature is raised to 5000$\K$ at 3.6$\gcc$, the number of
molecules decreases to 79\%, leading to 7\% atoms, 10\% O$_3$, and 4\%
O$_4$-like complexes. The lifetime of molecules decreases to about three
O$_2$ vibrations, indicating the presence of a dynamical dissociation
equilibrium with molecules still being the predominant and most stable
species.
At $\rho$=4.5$\gcc$ and T=6400$\K$, we find 9\% atoms and 58\%
molecules and also 16\% O$_3$ clusters. The remaining 17\% of the
nuclei belong to short-lived complexes composed of four or more atoms.
Further inspection reveals that O$_4$-like clusters arrange
predominantly into short zig-zag chains, consistent with the finding
of Ref.~\cite{NA02} for the compressed solid. At this high density,
the O$_2$ lifetime is only about 1.5 vibrational periods.  The short
lifetime of O$_3$ at all densities clearly indicates that triatomic
configurations are closer to scattering states than to stable chemical
species. Nevertheless, it is interesting to note that the ``bond''
angles of O$_3$ clusters are not randomly distributed but vary between
80$^\circ$ and 150$^\circ$ with an average of 120$^\circ$. Therefore,
the transient structure of O$_3$ clusters resembles that of the ozone
molecule.

\begin{figure}[htb]
\includegraphics[angle=0,width=\figurewidth]{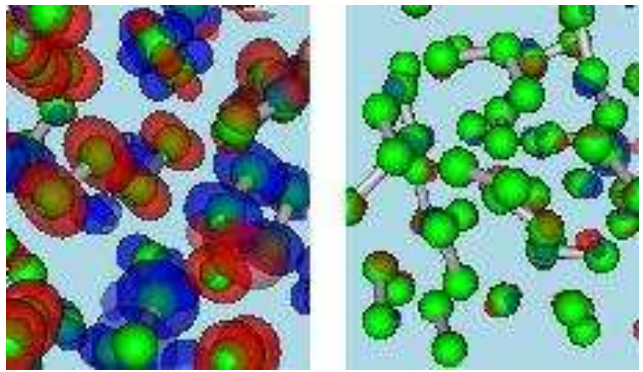}
\caption{Difference between spin-up and spin-down densities for two 
         representative snapshots at $\rho = 2.8\gcc$ and $T=1000\K$
         (left panel) and $\rho = 4.5\gcc$ and $T=6400\K$ (right
         panel). Positive and negative regions are represented by red
         and blue isosurfaces. Green spheres indicate the positions of
         the atoms.}
\label{spinfluctpicture}
\end{figure}

\begin{figure}[htb]
\includegraphics[angle=0,width=\figurewidth]{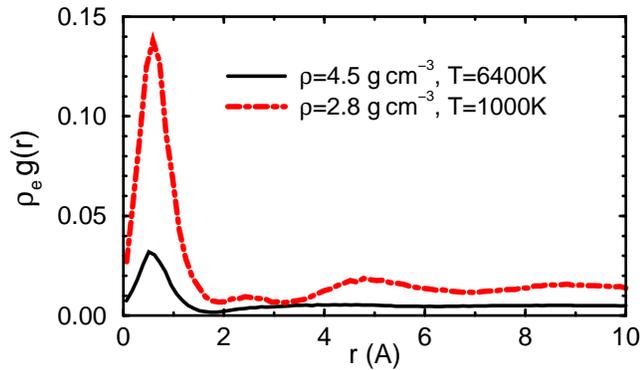}
\caption{Pair correlation function derived from nuclear positions and
         the absolute difference of the spin density. The first peak
         quantitatively characterizes spin fluctuations present in the
         molecular liquid at low density of 2.8$\gcc$. These
         fluctuations are suppressed at higher densities.}
\label{spinfluct}
\end{figure}

As mentioned above, while spin effects are negligible for the
structural properties of the fluid, it is crucial to include them
explicitly in order to understand the electronic properties of liquid
oxygen. In the low density regime, the liquid is composed of a
disordered arrangement of O$_2$ molecules carrying a net spin of one
as in the gas phase. The left panel of Fig.~\ref{spinfluctpicture}
shows the difference ($\Delta \rho_{\downarrow \uparrow}$) between
spin-up ($\rho_\uparrow$) and spin-down densities ($\rho_\downarrow$),
for a snapshot at $\rho=2.8\gcc$ and 1000$\K$, showing a random
distribution of up and down spins. At higher P (right panel of
Fig.~\ref{spinfluctpicture}), $\Delta \rho_{\downarrow \uparrow}$ has
almost vanished throughout the whole system, indicating that spin
fluctuations are suppressed. The effect shown qualitatively in
Fig.~\ref{spinfluctpicture} can be characterized quantitatively by the
correlation function of the nuclei $\rr_i$ and the electronic magnetic
moment, $m_z(\rr) = \mu_B[\rho_\uparrow(\rr)-\rho_\downarrow(\rr)]$,
\begin{equation}
\rho_e \, g_{{\rm O},m_z} (|\rr|) = \frac{1}{N} \left< \sum_{i=1}^{N} m_z(\rr_i-\rr) \right>
\quad,
\end{equation}
where $\rho_e$ is the electron density, $N$ the number of atoms and
$\left< \ldots \right>$ denotes an average over ionic configurations
and the solid angle of $\rr$. This correlation function is similar to
the nuclei-electron pair correlation function. The first, intense peak
in the low density correlation function shown in Fig.~\ref{spinfluct}
indicates that the majority of molecules are in the triplet state. At
higher density, the intensity of the first peak is decreased by almost
an order of magnitude, showing that most of the pairs in the fluid do
not carry a net spin. 

In order to qualitatively understand the effect of pressure on the
spin state of the oxygen molecule, we carried out calculations of the
total energies of O$_2$ in the singlet ($E_{S=0}$) and triplet
($E_{S=1}$) state in the presence of a confining potential mimicking an
average pressure effect. Interestingly, we found that the difference
$E_{S=0}-E_{S=1}$ can be substantially reduced by applying an external
confining potential.  This difference decreases as a function of the
confinement strength.

In order to compare our results with recent shock wave data, we
computed the conductivity ($\sigma$) of the liquid for selected
snapshots taken from our GCSDA simulations at different pressures,
using the Kubo-Greenwood formula~\cite{Si99,collins01}. We note that
in the low pressure regime, where spin fluctuations are present, GCSDA
and GGA results are very different. In particular, the GGA yields much
smaller optical gaps, consistent with the incorrect treatment of
excitation in the isolated molecule if spin is not explicitly taken
into account.  On the other hand, at higher densities, the computed
GGA and GCSDA conductivities are very similar.

\begin{figure}[htb]
\includegraphics[angle=0,width=\figurewidth]{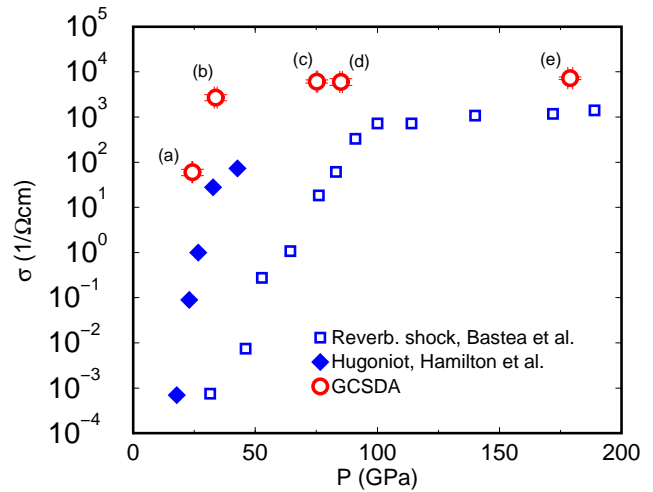}
\caption{Measured and calculated conductivity as a function of pressure. 
         The labels on the simulation data points correspond to the
         thermodynamic conditions specified in Fig.~\ref{gofr}.}
\label{cond}
\end{figure}

Our results are reported in Fig.~\ref{cond} along with the
experimental results of Ref.~\cite{Bastea01} and \cite{Ha88}. The
comparison between theory and experiment is not straightforward: the
temperature in the measurements by of Hamilton \etal \cite{Ha88} is in
general different from that of Bastea {\em et al.}'s
data~\cite{Bastea01}, and in both experiments, T was not measured but
estimated on the basis of empirical equation of state (EOS)
models. However, the two sets of experimental data show the same
qualitative rise of $\sigma$ as a function of P and T. In particular,
Bastea {\em et al.}'s data show a six order of magnitude rise in
conductivity with increasing shock compression, corresponding to a P
and T increase from about 30 to 120 GPa and $1000$ to $5000\K$,
respectively.  Our calculations within GCSDA show only two orders of
magnitude increase in $\sigma$ in the same P and T range.  This
increase of $\sigma$ as a function of $T$ occurs primarily in a state
where the fluid is entirely composed of stable O$_2$ molecules ($\rho
= 2.8 \gcc, T = 1000 \rightarrow 3000\K$). This indicates that in
oxygen, metallization is not accompanied by molecular dissociation as
found in other first row molecular fluids, in particular
hydrogen~\cite{collins01} and nitrogen~\cite{mazevet03}.
Above 100 GPa both theory and experiment show an almost constant $\sigma$
as a function of compression, but the theoretical values
are about a factor of 7 larger than the experimental ones.  

In our calculations, overestimates of conductivity values could be due
to the use of the GCSDA, which is known to underestimate electronic
gaps in many systems.  In order to qualitatively investigate this
effect, we used a scissors operator and arbitrarily increased the
value of the computed gaps by a factor of two. This lowered the
computed conductivity by one order of magnitude at $\rho$ = 2.8$\gcc$,
both at 1000 and 3000 K. However, even after this ``gap correction'',
our theoretical data show values of $\sigma$ approaching those of a
poor metal at a pressure much lower (30--40 GPa) than found
experimentally. The computed conductivity is then slowly changing as a
function of P, and it reaches metallic-like values at approximately 80
GPa. This ``slow'' metallization of the fluid as a function of
pressure may be related to spin fluctuations and their ``slow''
suppression as a function of pressure. At low P, the molecular fluid,
with each oxygen molecule in a triplet state, can be viewed as a
disordered spin system. As the pressure is increased, this system
shows a glass-like behavior due to the presence of a multitude of
local minima in its potential energy surface, corresponding to a
multitude of different spin arrangements.  Eventually, the fluid
reaches metallization when spin fluctuations are almost completely
suppressed.  In this respect, compressed liquid oxygen is unique among
first row molecular fluids and rather different from, e.g. liquid
hydrogen and liquid nitrogen, composed of molecules with spin zero.

In summary, our first principles simulations show that above 30 GPa,
over a wide pressure range of about 50 GPa, liquid oxygen transforms
from a semi-conducting to a metallic fluid by closure of the band gap
in a disordered molecular state. For $P \geq$ 80 GPa, molecular
dissociation is observed, giving rise to short-lived atomic- and
ozone-like species. We did not observe any stable O$_4$ molecule for
the thermodynamic conditions considered here, which is consistent with
recent theoretical findings for solid oxygen~\cite{NA02}. Our results
also show that spin fluctuations are suppressed at high pressure, with
no significant magnetic moment present in the high pressure liquid, as
opposed to the semi-conducting low pressure regime.

Our findings for the electrical conductivity point at the need for
further experimental and theoretical work. Experimentally, in shock
wave experiments only P and $\sigma$ are measured.  Data are then
analyzed using a model EOS as input for hydrodynamic codes which
compute densities. Finally, the temperature is inferred from
additional empirical chemical model calculations. This type of
analysis may introduce uncertainties in the estimated $\rho$ and T at
a given P and $\sigma$. Measurements of the oxygen EOS obtained with
shock wave experiments and possibly with other techniques, e.g. DAC
measurements, may help clarify the existing discrepancies between
theory and experiment.  Finally, we note that theoretical inaccuracies
on calculated conductivity values may arise from the use of an
approximate exchange and correlation functional. Although
computationally still out of reach, quantum Monte Carlo calculations
for liquid oxygen under pressure might be used in the future to
address this issue~\cite{HG02}.

\begin{acknowledgments}
We thank M. Bastea for her support of this project and for many useful
discussions, D.~Young for providing us with EOS tables from chemical
models, and L.~Benedict, V.~Recoules, E.~Schwegler, and T.~Ogitsu for
useful suggestions. This work was performed under the auspices of the
U.S. Dept. of Energy at the University of California/LLNL under
contract no. W-7405-Eng-48.

\end{acknowledgments}

\end{document}